# A Driver Behavior Modeling Structure Based on Non-parametric Bayesian Stochastic Hybrid Architecture

Hossein Nourkhiz Mahjoub
Networked Systems Lab
ECE Department
University of Central Florida
Orlando, FL, US
hnmahjoub@knights.ucf.edu

Behrad Toghi
Networked Systems Lab
ECE Department
University of Central Florida
Orlando, FL, US
toghi@knights.ucf.edu

Yaser P. Fallah
Networked Systems Lab
ECE Department
University of Central Florida
Orlando, FL, US
yaser.fallah@ucf.edu

*Abstract*—Heterogeneous nature of the vehicular networks, which results from the co-existence of human-driven, semi-automated, and fully autonomous vehicles, is a challenging phenomenon toward the realization of the intelligent transportation systems with an acceptable level of safety, comfort, and efficiency. Safety applications highly suffer from communication resource limitations, specifically in dense and congested vehicular networks. The idea of model-based communication (MBC) has been recently proposed to address this issue. In this work, we propose Gaussian Process based Stochastic Hybrid System with Cumulative Relevant History (CRH-GP-SHS) framework, which is a hierarchical stochastic hybrid modeling structure, built upon a non-parametric Bayesian inference method, i.e. Gaussian processes. This framework is proposed in order to be employed within the MBC context to jointly model driver/vehicle behavior as a stochastic object. Non-parametric Bayesian methods relieve the limitations imposed by non-evolutionary model structures and enable the proposed framework to properly capture different stochastic behaviors. The performance of the proposed CRH-GP-SHS framework at the inter-mode level has been evaluated over a set of realistic lane change maneuvers from NGSIM-US101 dataset. The results show a noticeable performance improvement for GP in comparison to the baseline constant speed model, specifically in critical situations such as highly congested networks. Moreover, an augmented model has also been proposed which is a composition of GP and constant speed models and capable of capturing the driver behavior under various network reliability conditions.

## I. INTRODUCTION

Individual vehicular network agents are anticipated to perform wiser cooperative decisions if they continuously be acquainted with either the exact or the most probable actions of other agents within a time horizon ahead. This imperative notion, which is usually referred to as *situational awareness* in the context of vehicular networks, is mainly achieved by the virtue of sensory information, captured locally through sensors such as radar, LiDAR, or camera, along with the communicated information between the network elements. Communication-driven portion of the situational awareness data is crucial due to the inherent restrictions of sensory information, such as obstacle blocking or environmental issues (fog, rain, dimness, etc.).

However, the limited communication bandwidth of the currently available vehicular communication standards, such as Dedicated Short Range Communication (DSRC) [1], strongly motivates the development of optimized schemes for disseminating critical information among vehicles within a certain required vicinity. The minimum required robust communication range is mainly forced by the application layer. More specifically, cooperative safety applications, such as Forward Collision Warning/Avoidance (FCW/A) [2]-[4], Cooperative Adaptive Cruise Control (CACC) [5], Lane-Keep-Assistance (LKA) [6], etc., which are supposed to employ the achieved situational awareness to make proper safety and efficiency decisions, play the main role to determine this range. According to technical documents from National Highway Traffic Safety Administration (NHTSA), 300m distance is considered as the minimum required range for a generic V2V communication standard, such as DSRC [7]. Therefore, the bandwidth limitation challenge within this proximity should be appropriately handled to guarantee the robust reception of neighboring vehicles' information.

Several DSRC-based congestion control methods have been proposed in the literature [8]-[16] aim to optimize the Basic Safety Message (BSM) formation and communication strategy through continuous acclimation of different flexible DSRC standard parameters, such as transmission power, rate, and message content, to the network conditions. The communication channel utilization efficiency has been notably improved using aforementioned adaptive mechanisms and some of them, such as [16], have been selected as the core congestion control algorithms of the SAE J2945/1 standard [17]. However, all these mechanisms are developed assuming the fact that the broadcast message should be finally filled by raw information, directly sourced from CAN-bus and GPS.

An innovative idea has been recently proposed in [18] and more investigated in [19] as the model-based communication (MBC) which proposes a new design perspective to be utilized for the DSRC congestion control problem. This methodology proposes to replace the mixed vehicle/driver behavior with an abstract description (model) and then share the models and their updates over the network instead of directly communicating raw dynamic information. Since this approach shifts the paradigm and changes the solution domain from any scheme of raw data communication to model based information networking, it

This material is based on work supported by the National Science Foundation under CAREER Grant 1664968.

seems very promising towards a notable enhancement in channel utilization performance. The other significant advantage of MBC methodology against raw data communication schemes, in addition to its potentially higher channel utilization capability, is its capability to substantially increase the forecasting accuracies over longer prediction horizons. This is due to the flexibility of this approach to update the model structure and/or parameters at the host vehicle, the vehicle that generates the model, and then update the remote vehicles' knowledge of the updated models on the fly.

Proper model derivation strategies which are capable of capturing the high level (long-term) driving behaviors, i.e. driving maneuvers, while can simultaneously follow the low-level (short-term) dynamic trends within each maneuver could profoundly outperform the conventional forecasting schemes at the remote (receiver) vehicles. It is worth mentioning that a precise MBC-customized communication policy is also essential in conjunction with this framework to accomplish the above mentioned goals. In the conventional framework, which is currently the dominant adopted strategy in the vehicular industry, remote vehicles always assume a predefined behavior (or roughly speaking a predefined model) of the host vehicle, e.g. constant acceleration or constant speed model, with no structural model evolutions over time. However, the composition of the complex VANETs, which is normally a mixture of human driven, semi-automated and fully autonomous nodes, forms a highly dynamic network and imposes a high level of stochasticity in the model structures. Therefore, the weak predefined model structure assumption which neglects the plausible model structural evolutions over time, results in a poor prediction quality. This precision deficiency, inspires the development of more advanced modeling schemes which are capable of capturing the model evolution trends and handle the modeling task more rigorously. This is an inevitable requirement to realize a trustable situational awareness in contemporary vehicular networks.

In this work, we have proposed a hybrid stochastic modeling framework within the MBC context which is built upon a powerful Bayesian non-parametric inference scheme, i.e. Gaussian Processes. The proposed framework, which has been depicted in Figure 1 and will be referred to as Gaussian Process-Stochastic Hybrid System with Cumulative Relevant History (CRH-GP-SHS) from now on, combines the impressive flexibility and forecasting capabilities of non-parametric Bayesian methods with the apprehensible SHS modeling procedure and tries to increase the resultant model precision using a high-level online maneuver-based training history selection scheme. Performance of the proposed framework, at inter-mode level, has been investigated for a specific maneuver, i.e. lane change, to demonstrate its performance enhancement capability against the state of the art prediction methods currently utilized in realistic vehicular industry. Constant speed model is selected in this work as the baseline for comparison and performance evaluation. The analysis is performed on a set of real lane change maneuvers from NGSIM-US101 [20], which is a realistic dataset has been available by US-DoT.

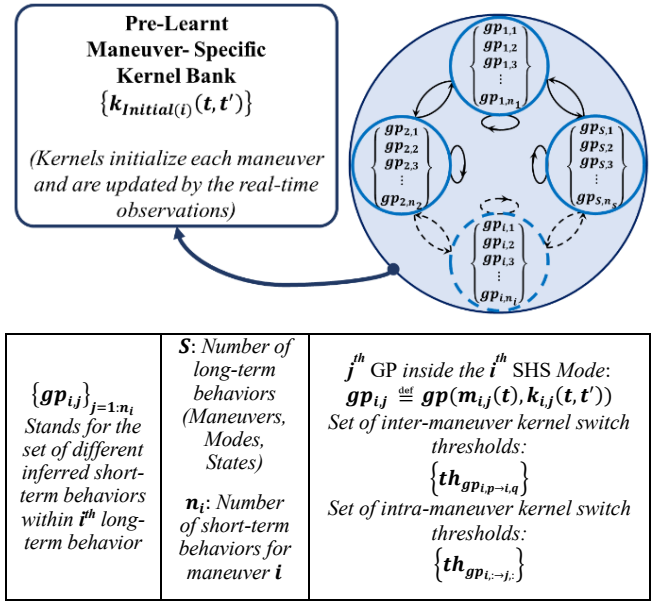

Figure 1. Proposed Gaussian Process-Stochastic Hybrid System with Cumulative Relevant History (CRH-GP-SHS) Framework

The rest of this paper is organized as follows. Section II is devoted to the targeted problem statement, in addition to a brief theoretical explanation of the underlying building blocks of the proposed framework, i.e. non-parametric Bayesian Gaussian process inference, and stochastic hybrid systems notion. The evaluation details are presented in section III. Finally, we conclude the paper with a summary in section IV.

## II. PROBLEM STATEMENT

### A. Gaussian Process Based Stochastic Hybrid System with Cumulative Relevant History (CRH-GP-SHS) Framework

As mentioned earlier, the final goal of the modeling framework investigated in this paper, i.e. CRH-GP-SHS framework, is deriving precise predictive models for both short-term and long-term mixed driver/vehicle behaviors. trends (almost within 0-3 seconds) of the critical vehicle dynamic states, such as its position, velocity, acceleration, etc., inside the framework discrete modes which are equivalent to different long-term behaviors (maneuvers) of the driver. The proposed approach to fulfill this goal is building a cumulative maneuver-specific training history on the fly from the identical or relevant observed maneuvers in recent driving history of the driver, and then feeding this training data to the model inference block, i.e. Gaussian process block, as its initial training set. This initial training set will be updated in an online manner by adding new observations from the currently ongoing maneuver in order to force the inference procedure to consider the current cognitive state and actions of the driver in its model derivation procedure. When each long-term maneuver is finished, its data is added to the training data bank of that specific maneuver to be used as part of the initial training history for the next similar maneuver. Another equivalent approach to the aforementioned method of initial training set augmentation by new observed relevant maneuvers, is creating a maneuver-specific model bank from the currently observed relevant maneuvers and then feed this pre-learned model parameter as the initial parameter values for model inference in conjunction with the ongoing maneuver data as the model inference training set. This second method, tries to fine-tune the pre-learned model of this maneuver and adapt its inferred parameters to the driver's current behavior, which might come from a different cognitive status, such as distraction, haste, drowsiness, etc. These two approaches are theoretically equivalent, but the second one should be more appropriate for online applications, such as our application, since it already has a pre-learned model available and needs to process this model using a short observation set, coming from the maneuver which is currently in progress. Theoretical aspects of the short-term behavior inference method, i.e. Gaussian process regression, is briefly covered in the following sub-section.

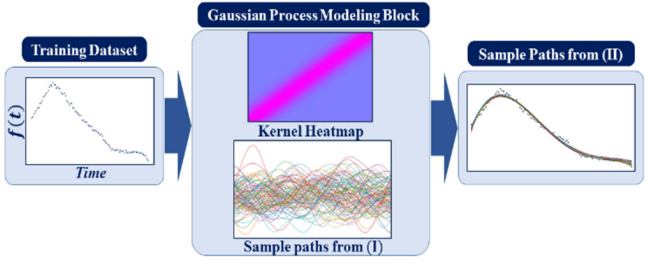

Figure 2. General representation of the Gaussian process inference of function *f(t)* conditioned on a noisy set of observations. (I) Prior (II) Posterior distribution over the underlying functions

### B. Non-Parametric Bayesian Inference based on Gaussian Process Regression

Different vehicle dynamic states which are intended to be modeled based on our earlier discussion in section II.A, could be considered as individual time-series which should be regressed using an appropriate regression method. In this framework, targeted time-series values, e.g., position values, are assumed to be generated by some underlying function of time, $f(t)$. Therefore, the regression problem here is equivalent to discovering the characteristics of this function through a finite set of its available realizations. We propose to use the Bayesian inference framework here, focusing on non-parametric regression. The goal in a non-parametric regression method is to find the best function representation for the observed time-series values without any prior assumption on the form of the underlying function. We plan to use a non-parametric Bayesian inference method, i.e., Gaussian process regression, to derive the model of the host vehicle and its driver as a unique object. The outcome is a set of functions describing the underlying modes that generate the behavior of the driver/vehicle combination.

Gaussian process regression is powerful class continuous function modeling. Gaussian processes could also be utilized for classification purposes, instead of regression, in the case of discrete function values, which is out of this paper's scope. A formal definition of Gaussian process could be represented as follows [21]:

A Gaussian process defines a distribution over function values $f(t)$ at any arbitrary point within the function input range, such that any finite subset of the drawn function values from this distribution form a multivariate Gaussian random vector (have joint Gaussian distribution). The common notation for this definition is as follows:

$$f(t) \sim gp(m(t), k(t, t')) \quad (1)$$

$$\{X_i\}_{i=1,2,...,m} = \{f(t_i)\}_{i=1,2,...,m} \sim N(\bar{\mu}, \Sigma) \quad (2)$$

$$\bar{\mu} = m(t_i); \ \Sigma_{i,j} = k(t_i, t_j) \quad \forall i, j \in \{1, 2, ..., m\} \quad (3)$$

A general pictorial representation of the sample paths from a Gaussian process prior and posterior distribution is depicted in Figure 2. Posterior distribution is inferred by conditioning the problem on a set of noisy observations as the training data.

Gaussian process regression model assumes each observed value as a draw from a normal random variable. Therefore, the set of $m$ observations form an $m$-dimensional multivariate normal random vector. This multivariate random vector is defined by a mean vector of length $m$ plus an $m \times m$ covariance matrix. This covariance matrix is called the kernel in the Gaussian process context. This $m$-dimensional multivariate (jointly) normal random vector is considered as the marginal distribution of the underlying infinite dimensional vector, i.e., underlying function after integrating out the function values at all unobserved input points. Kernel matrix defines the correlation between the elements of this $m$-dimensional marginal distribution. It is notable that the whole modeling innovation of the GP method is almost conveyed by the kernel function characteristics, rather than the mean function. Different types of kernels have been proposed in the literature, such as linear, squared exponential, and spectral mixture, to name a few. Each kernel type is appropriate to capture a specific pattern in the time-series evolution. Due to the nature of the problem which has been tackled in this work, we picked the linear kernel for our modeling purposes. More specifically, since the baseline information recording time intervals, i.e. 100 ms, is in the order of magnitudes of our mechanical system (vehicle/driver) response time (due to the internal physical system inertia), high rate fluctuations cannot usually be observed in the system physical dynamics records. Since linear kernel is not only a function of difference between input values, and consequently is not invariant with respect to the input domain translations, it is categorized under non-stationary kernel types. This kernel could be formulated as follows:

$$k(t, t') = \sigma_L^2 (t - c_1)(t' - c_2) \quad (4)$$

### III. MODEL EVALUATION AND ANALYSIS

In order to evaluate the performance of the proposed method, the aforementioned Gaussian process-based modeling scheme has been applied on an extensive set of realistic driving maneuvers available in the NGSIM-US101 dataset. More specifically, our focus in this work is on a discrete mode (a maneuver) of the CRH-GP-SHS framework, in order to demonstrate the effectiveness of GPs for short-term pattern modeling within the system discrete modes. The lane change maneuver has been chosen in this work as a specific long-term driver behavior, and we tried to model the lateral position of the vehicle, as one of the most critical dynamic parameters required by safety applications in lane change scenarios, through the available set of its already observed instances. It should be emphasized that these position values are basically function of both the mechanical plant (vehicle) and the stochastic human-driven actions taken by the driver. Thus, the aforementioned problem can be noted as modeling a stochastic system, which has several latent variables, e.g. driver cognitive state, weather and road situation, etc., which are not directly available from the dataset. This point justifies the higher performance of a stochastic modeling framework, instead of using certain predefined models without any structural evolution over time.

It is assumed here that, in the case of our problem, the host vehicle desires to model its own future behavior through its own available history recorded with 10 Hz rate. These models can be communicated afterwards over the network and be utilized in other (remote) vehicles in order to enable them to predict the host vehicle's behavior, without reception of new raw information or model updates during a certain time period, referred as the forecast horizon. Therefore, we tried to evaluate the precision of the achieved forecasted values by the model over the forecast horizons ranging from 100ms (1 sample) to 3 seconds (30 samples) after the last observation instance. 40 lane change maneuvers have been chosen from the NGSIM-US101 dataset for our analysis purposes. Each maneuver has a 3 second duration and is symmetrically truncated from the rest of the dataset, i.e. from 1.5 seconds before the reported lane change moment in the dataset to 1.5 second after that. The lateral displacement has also been cross validated in order to confirm the existence of a true and complete lane change within the mentioned duration for each selected maneuver. For all of the selected maneuvers, almost 10 feet of lateral displacement has been observed, which is identical to the typical lane width and confirms the correctness of the recorded lane change trajectory.

The following conclusions could be derived out of the results:

- GP regression modeling scheme strongly outperforms the baseline model for forecast horizon values greater than 1 second ahead, which could be equivalently interpreted as the reception rates under 1Hz (Figure 3). In this figure the relation between the forecast horizon (in seconds) and the reception rate (in Hz) is as follows:

$$Reception\ Rate\ (Hz) = \frac{1}{(Forecast\ Horizon\ (s) + 0.1)} \quad (5)$$

Therefore, by sweeping the forecast horizon from 100ms to 3s (or equivalently from 1 to 30 samples ahead), the reception rate sweeps from 5 Hz to almost 0.32 Hz.

Considering the discussed observations, it can be concluded that for highly congested network situations, if the baseline model be augmented by the GP regression model with a linear kernel, the resultant compound model will be capable of forecasting both near (less than 1 seconds ahead) and far future (between 1 and 3 seconds ahead) cases. In this compound model, the baseline sub-model is responsible for the former prediction duration, while GP sub-model handles the latter. It is worth mentioning that far future prediction accuracy is essential in congested networks where the model updates cannot be received frequently by the remote vehicles.

- The far future behavior prediction accuracy at the starting moments of the maneuver is highly essential for the controllers which handle the remote vehicles' safety applications' tasks. A higher accuracy prediction for the farther future instances provides more adequate models to these controllers for a longer future horizon, which in turn increases their capability of making smoother and wiser control actions. This statement is very important, especially if the safety applications employ model predictive controllers (MPC) to achieve their goals. Therefore, we have also evaluated the far future accuracy prediction of the GP model versus baseline in Figure 4., calculated over the whole set of selected 40 lane changes. This figure, which shows a notable higher performance for GP in this case, presents the 95% of the absolute error values over far prediction horizons (2-3 seconds ahead) when the lane change maneuver is in its beginning phase (during first 1s of the lane change maneuver). Horizontal axis in this plot sweeps the number of consecutive observed relevant maneuvers, i.e. lane changes, which have been augmented in the training history one by one. This number ranges from 1 to 40 lane changes in our analysis. So, one could interpret this figure in the following way: assume a driver starts a trip and whenever he makes a complete relevant maneuver during this trip, this maneuver is added to the training history which will be used for the next similar maneuver. So the training history becomes richer gradually by augmenting more and more relevant data when the driver continues his trip. It could be concluded that although increasing the number of augmented relevant maneuvers in the training history does not necessarily reduce the prediction error, GP method accuracy always dominates the baseline model.

- The overall performance of the GP model versus the baseline, over the whole range of forecast horizon (0-3 seconds) and during the complete lane change maneuver has been evaluated for all 40 lane change maneuvers. For this analysis, same as Figure 4, these 40 maneuvers have been sorted in a sequence from lane change 1 to lane change 40 where the $i^{th}$ lane change in the sequence is assumed to have the previous $(i-1)$ lane changes as relevant maneuvers in its training history. The results, which are depicted in Figure 5, show the noticeable higher accuracy for GP versus the baseline.

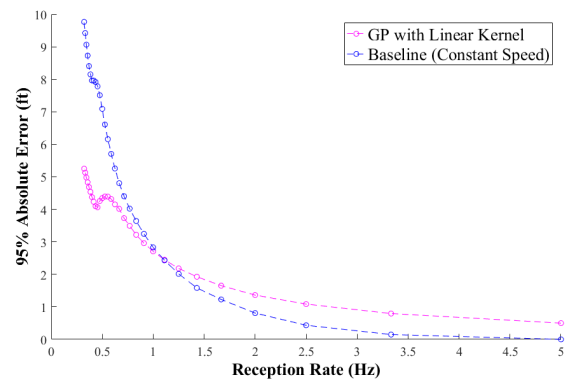

Figure 3. 95% absolute error vs. reception rate for a complete lane change maneuver (3 seconds)

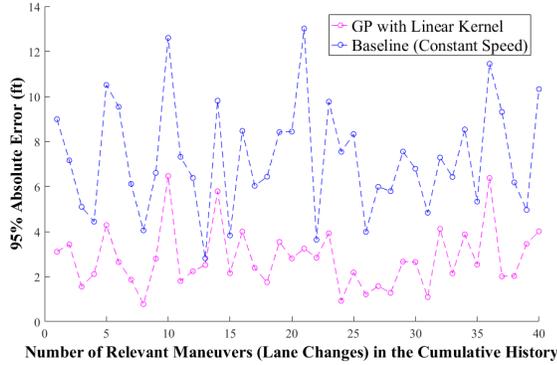

Figure 4. 95% absolute error for 2-3 seconds ahead prediction horizon during the starting phase of the lane change maneuver (first 1s of the maneuver) calculated over all maneuvers.

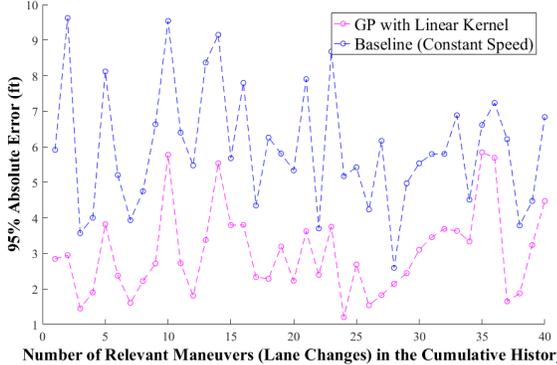

Figure 5. 95% absolute error for 0-3 seconds ahead prediction horizon during the complete lane change maneuver, calculated over all maneuvers.

## IV. CONCLUDING REMARKS

In this work CRH-GP-SHS framework, which is a stochastic hybrid system modeling framework based on a non-parametric Bayesian regression scheme, i.e. Gaussian processes, has been proposed and evaluated. This framework could be employed within the context of model-based communication to jointly model the vehicle/driver behavior through a set of observed relevant maneuvers in the recent history of the driver. This framework has been evaluated at the inter-mode level here. The inter-mode level is responsible for capturing the short-term behavioral evolutions over time (within around 3-5 seconds ahead) during one specific long-term behavior (maneuver or mode). The performance of the proposed approach to track the joint vehicle-driver behavior is investigated via its ability to forecast the position of the subject vehicle. Lane change has been selected as the long-term behavior here and a set of lane change maneuvers from NGSIM-US101 realistic dataset has been employed for this analysis. The results show a significant tracking precision improvement against the constant speed model, which is one of the state of the art prediction methods in the current vehicular industry, as the baseline.

Deriving similar models for different other maneuvers, such as free following, take over, joining and leaving a platoon, etc., and then merging them together to form a complete CRH-GP-SHS model is the future research directions which are now under investigation by the authors.

REFERENCES

[1] J. B. Kenney, "Dedicated Short-Range Communications (DSRC) Standards in the United States," in *Proceedings of the IEEE*, vol. 99, no. 7, pp. 1162-1182, July 2011.
[2] J. Wang, C. Yu, S. E. Li and L. Wang, "A Forward Collision Warning Algorithm with Adaptation to Driver Behaviors," in *IEEE Transactions on Intelligent Transportation Systems*, vol. 17, no. 4, pp. 1157-1167, April 2016.
[3] S. M. Iranmanesh, H. N. Mahjoub, H. Kazemi and Y. P. Fallah, "An Adaptive Forward Collision Warning Framework Design Based on Driver Distraction," in *IEEE Transactions on Intelligent Transportation Systems*.
[4] F. Muehlfeld, I. Doric, R. Ertlmeier and T. Brandmeier, "Statistical Behavior Modeling for Driver-Adaptive Precrash Systems," in *IEEE Transactions on Intelligent Transportation Systems*, vol. 14, no. 4, pp. 1764-1772, Dec. 2013.
[5] H. Kazemi, H. N. Mahjoub, A. Tahmasbi-Sarvestani, and Y. P. Fallah, " A Learning-based Stochastic MPC Design for Cooperative Adaptive Cruise Control to Handle Interfering Vehicles," to be appeared in *IEEE Transactions on Intelligent Vehicles*.
[6] F. Breyer, C. Blaschke, B. Farber, J. Freyer and R. Limbacher, "Negative Behavioral Adaptation to Lane-Keeping Assistance Systems," in *IEEE Intelligent Transportation Systems Magazine*, vol. 2, no. 2, pp. 21-32, Summer 2010.
[7] "Preliminary Regulatory Impact Analysis FMVSS No. 150, Vehicle-to-Vehicle Communication Technology for Light Vehicles," NHTSA DOT HS 812 359, Dec. 2016.
[8] Y.P. Fallah, C.L. Huang, R. Sengupta, and H. Krishnan, "Analysis of information dissemination in vehicular ad-hoc networks of cooperative vehicle safety systems," *IEEE Trans. on Vehicular Technology*, vol. 60, no. 1, pp. 233–247, January 2011.
[9] Y.P. Fallah, C.L. Huang, R. Sengupta, and H. Krishnan, "Congestion control based on channel occupancy in vehicular broadcast networks," IEEE Vehicular Technology Conference (VTC-Fall), September 2010
[10] G. Bansal, H. Lu, J. Kenney, and C. Poellabauer, "EMBARC: Error model based adaptive rate control for vehicle-to-vehicle communications," Proc. ACM International Workshop on Vehicular Inter-Networking, Systems, Applications (VANET), June 2013, pp. 41–50
[11] C.L. Huang, Y.P. Fallah, R. Sengupta, and H. Krishnan, "Adaptive intervehicle communication control for cooperative safety systems," *IEEE Network*, vol. 24, Issue 1, pp. 6–13, January-February 2010.
[12] G. Bansal, J. Kenney, and C. Rohrs, "LIMERIC: A linear adaptive message rate algorithm for DSRC congestion control," *IEEE Trans. Vehicular Technology*, vol. 62, no. 9, pp. 4182–4197, November 2013.
[13] H. Lu, G. Bansal, and J. Kenney, "A joint rate-power control algorithm for vehicular safety communications," Proc. ITS world congress 2015.
[14] A. Weinfied, J. Kenney, and G. Bansal, "An adaptive DSRC message transmission interval control algorithm," Proc. ITS World Congress, October 2011, pp. 1–12.
[15] S. M. O. Gani, Y. P. Fallah, G. Bansal and T. Shimizu, "A Study of the Effectiveness of Message Content, Length, and Rate Control for Improving Map Accuracy in Automated Driving Systems," in IEEE Transactions on Intelligent Transportation Systems.
[16] Y. P. Fallah, N. Nasiriani and H. Krishnan, "Stable and Fair Power Control in Vehicle Safety Networks," in *IEEE Transactions on Vehicular Technology*, vol. 65, no. 3, pp. 1662-1675, March 2016.
[17] SAE International, "Surface Vehicle Standard – On-Board System Requirements for V2V Safety Communications," J2945TM/1, Issued 2016-03.
[18] Y. P. Fallah, "A model-based communication approach for distributed and connected vehicle safety systems," *2016 Annual IEEE Systems Conference (SysCon)*, Orlando, FL, 2016, pp. 1-6.
[19] E. Moradi-Pari, H. N. Mahjoub, H. Kazemi, Y. P. Fallah and A. Tahmasbi-Sarvestani, "Utilizing Model-Based Communication and Control for Cooperative Automated Vehicle Applications," in *IEEE Transactions on Intelligent Vehicles*, vol. 2, no. 1, pp. 38-51, March 2017.
[20] NGSIM Homepage [Online] Available: https://www.fhwa.dot.gov/publications/research/operations/07030/
[21] Gaussian Processes for Machine Learning Carl Edward Rasmussen and Christopher K. I. Williams The MIT Press, 2006. ISBN 0-262-18253-X.